\documentclass[useAMS,usenatbib]{mn2e}

\usepackage{graphicx}
\usepackage{amssymb}
\usepackage{color}
\usepackage{pst-grad}
\usepackage{psfrag}

\newcommand{\pdiff}[2]{\frac{\partial #1}{\partial #2}}
\newcommand{\dif}[2]{\frac{{\rm d} #1}{{\rm d} #2}}

\newcommand{\eqb}{\begin{eqnarray}}
\newcommand{\eqe}{\end{eqnarray}}
\newcommand{\diff}{{\rm d}}

\newcommand{\n}{\noindent}
\newcommand{\bfm}[1]{\mathbf{#1}}
\newcommand{\beq}{\begin{equation}}
\newcommand{\eeq}{\end{equation}}

\newcommand{\gesim}{\,\raisebox{-0.4ex}{$\stackrel{>}{\scriptstyle\sim}$}\,}
\newcommand{\lesim}{\,\raisebox{-0.4ex}{$\stackrel{<}{\scriptstyle\sim}$}\,}

\title[The transport of cosmic rays in self-excited magnetic turbulence]
{The transport of cosmic rays in self-excited magnetic turbulence}
\author[B. Reville, S. O'Sullivan, P. Duffy, J.G. Kirk]{B. Reville$^{1}$\thanks{E-mail:
brian.reville@mpi-hd.mpg.de}, S. O'Sullivan$^2$, P.
Duffy$^3$ and J.G. Kirk$^1$\\
$^1$Max-Planck-Institut f\"ur Kernphysik, Heidelberg 69029, Germany\\
$^2$School of Mathematical Sciences, Dublin City University, Glasnevin, Dublin 9, Ireland\\
$^3$UCD School of Physics, University College Dublin, Belfield, Dublin 4, Ireland}
\begin{document}

\date{Submitted December 3rd, 2007,....}

\pagerange{\pageref{firstpage}--\pageref{lastpage}} \pubyear{2007}

\maketitle

\label{firstpage}

\begin{abstract}
The process of diffusive shock acceleration relies on the efficacy 
with which hydromagnetic waves can scatter charged particles
in the precursor of a shock. The growth of self-generated waves
is driven by both resonant and non-resonant processes.
We perform high-resolution magnetohydrodynamic simulations of the 
non-resonant cosmic-ray driven instability, in which the unstable waves 
are excited beyond the linear regime.
In a snapshot of the resultant field, particle transport simulations are
carried out. The use of a static snapshot of the field is reasonable given that the Larmor period for particles is typically very short relative to the instability
 growth time. The diffusion rate is found to be close to, or below, the Bohm 
limit for a range of energies.
This provides the first explicit demonstration that self-excited turbulence
reduces the diffusion coefficient and has important implications for cosmic ray
transport and acceleration in supernova remnants.
\end{abstract}


\section{Introduction}
\label{sec-intro}

Supernova remnants (SNR) have long been identified as possible locations for the production
and acceleration of galactic cosmic rays. The diffusive shock acceleration 
mechanism provides a natural explanation for the observed power-law spectrum for these cosmic rays. 
Acceleration to high energies relies on confinement of
relativistic particles to the accelerating region close to the shock. Pitch-angle
scattering from self-produced hydromagnetic waves can provide a suitable mechanism,
but a detailed analysis demonstrates that the spectrum still falls
short of the cosmic ray knee at $\approx 10^{15}$ eV \citep{LagageCesarsky}. 
However, theory dictates that the maximum energy of the cosmic rays is determined by the diffusion coefficient, which is assumed to scale with the strength of
the ambient magnetic field. 
Therefore, in the quasilinear framework, 
amplification of the apparent ambient magnetic field in the vicinity
of the shock facilitates acceleration to higher energies.

To within an order of magnitude the timescale for the diffusive shock 
acceleration of a particle is
$ t_{\rm acc} \approx {\kappa}/{v_{\rm sh}^2}$
where $\kappa\equiv \lim_{t\to\infty}  \langle \Delta^2 r\rangle/2t$ is the asymptotic spatial diffusion coefficient where $\Delta r$ is the spatial displacement over time interval $t$. The conventional approximation for transport of particles of speed $v$ via scattering is Bohm diffusion which assumes a single scattering event occurs during each Larmor time. 

In the region upstream of a shock, the scattering of particles is due to 
resonant collisions with slowly moving magnetic irregularities which, in turn, are amplified by such interactions themselves. These turbulent irregularities,
until recently, have been assumed to saturate
at levels $\delta B \lesssim B_0$ \citep{mckenzievoelk82}. With such turbulence, Bohm diffusion can be
taken as a lower limit for the diffusion coefficient,  $\kappa \gesim \kappa_{\rm Bohm}$. In highly
turbulent fields, however, numerical investigations suggest that
particle transport properties may differ significantly
from this approximation \citep{Casseetal}.

\cite{belllucek} proposed a process of resonant amplification of the magnetic
field driven by the pressure gradient of cosmic rays, allowing rapid transfer
of energy into the Alfv\'en waves. 
Amplification factors of $\delta B / B_0 \sim 1000$ were estimated 
from the linear theory, 
although this value is possibly excessive as the analysis most likely
does not hold beyond $\delta B / B_0 \sim 1$. Nevertheless, 
a more detailed non-linear analysis does suggest acceleration up to  
$E_{\rm max} \sim 10^{16} {\rm eV}$. In particular, under expansion 
into a stellar wind, it may even be possible to 
achieve proton energies of the order $E_{\rm max} \sim 10^{17}$eV 
via this process.

Moving beyond this result, \cite{Bell2004} identified a non-resonant instability driven by streaming, positively charged
cosmic rays that operates on length scales much shorter than the Larmor radius $r_{\rm g}\equiv p/ e B_0$ of the driving particles (where $p$ is the particle momentum, $e$ is the electronic charge, and $B_0$ is the mean magnetic field magnitude). Crucially, under typical conditions in the precursor of a
high Alfv\'en-Mach-number supernova remnant shock, 
the growth rate for this non-resonant instability 
can be several orders of magnitude greater than that of 
its resonant counterpart. Indeed, numerical magnetohydrodynamic 
simulations carried out in the same work confirmed that field amplification by a factor of an order of magnitude is possible. 

Only recently have analytical approaches to particle transport 
achieved results consistent with numerical approaches. In the case of 
interplanetary shocks it has been shown that using detailed models of
particle transport in turbulent environments  
sub-Bohm diffusion can be achieved (see \cite{Zank} and references therein).

In this paper we investigate, using high resolution MHD simulations,
the non-resonant Bell-type instability and its non-linear evolution.
We also determine the effective transport properties
in the amplified field and compare them 
with the standard Bohm diffusion description.

The structure of the paper is as follows. In Section~\ref{sec-inst}, we 
recall the linear MHD analysis of the dispersion relation relevant
to the numerical simulations performed. 
In Section~\ref{sec-field} we describe the numerical method 
used to investigate the non-resonant instability and discuss some properties of the field produced. In Section~\ref{sec-part}, we describe the techniques used to integrate equations of motion in the snapshot field obtained from the 
simulations. We discuss the transport properties of both field lines and 
test particles. Finally, in Section~\ref{sec-conc}, we present our conclusions.

\section{The non-resonant instability}
\label{sec-inst}

Following \cite{Bell2004}, we determine the dispersion relation for waves in
the precursor of a quasi-parallel non-relativistic shock, where the energy density
of the cosmic rays is comparable to the ram pressure of the shock ($\sim 10\%$).
The usual assumptions made in the linear analysis of MHD waves propagating in a plasma are that the plasma is quasi-neutral, i.e. is charge neutral on aggregate. Therefore, assuming a plasma consisting entirely of protons and electrons in the presence of a pure proton cosmic ray component, the thermal plasma has a
charge excess of electrons. In the reference frame in which the upstream protons
are at rest, the cosmic rays are seen to stream with a group velocity 
approximately equal to the shock velocity, 
$\langle v \rangle \approx {v}_{\rm sh}$. 
A charge flux is induced in thermal plasma to neutralise
this current. In the case of a mean field $\bfm{B}_\|$ parallel to the shock
normal, this return current has a mean (unperturbed) value given by 
$\bfm{j}_\| = n_{cr} e \bfm{v}_{\rm sh}$ where $n_{cr}$ is the number density of cosmic rays and $e$ is the electronic charge. The momentum equation is given 
in the MHD approximation by

\begin{equation}
\label{MHDmom}
 \rho \frac{\diff {\bf u}}{\diff t} =
-\nabla P +{\bf j}_{\rm th}\times{\bf B} ,
\end{equation}

\n where ${\bf j}_{\rm th}$, the current density carried by the thermal plasma, is determined by Amp\`{e}re's law, 

\begin{equation}
 \nabla \times {\bf B} = \mu_0 \sum_s n_s q_s {\bf v}_s=
\mu_0({\bf j}_{\rm th}+{\bf j}_{\rm cr}) ,
\end{equation}
\n where $s$ is an index used to sum over the charged species and $\mu_0$ is the plasma permeability. 
The zeroth order cosmic-ray current density is parallel to the mean magnetic 
field, which is chosen to lie along the z-axis. 
All first order perturbations are taken to lie perpendicular to the 
zeroth order components such that the magnetic field and 
cosmic-ray current density can be expressed as 
$\bfm{B}=\bfm{B}_\|+\bfm{B}_\perp$, $\bfm{j}_{\rm cr}=\bfm{j}_\|+\bfm{j}_\perp$ respectively. 
Making use of the  induction equation

\begin{equation}
 \frac{\partial {\bf B}}{\partial t} = \nabla \times ({\bf u \times B})
\end{equation}

\n we look for plane-wave solutions to the linearised system with $\bfm{k}$ also parallel to the mean field. All
first order quantities $\bfm{B}_\perp$, $\bfm{j}_\perp$, $\bfm{u}_\perp$ and $\bfm{E}_\perp = -\bfm{u}_\perp\times\bfm{B}_\|$, take the circularly polarised form 
$A_x\pm iA_y = A_0 \exp[i(k z-\omega t)]$. For the purposes of analysis, the plasma is taken to be initially at rest. 

Using a kinetic description, it was shown by \cite{revilleetal07} 
that for modes with wavenumbers
$kr_{\rm g1}\gg 1$, where $r_{\rm g1}\equiv p_1/e | B_\| |$ is the Larmor radius of the cosmic rays of minimum momentum $p_1$, kinetic effects can be neglected
i.e. both thermal effects and perturbations to the mean 
cosmic-ray flux are negligible. Splitting the perpendicular
magnetic field perturbation into its cartesian components in the
x-y plane the relation 

\begin{eqnarray}
\left(\begin{array}{cc}
\omega^2-v_{\rm a}^2k^2 & i B_\| j_\| k/\rho\\
-i B_\| j_\| k/\rho & \omega^2-v_{\rm a}^2k^2
\end{array}\right) \left(\begin{array}{c}
 B_x \\ B_y
\end{array}\right) = 0
\label{eqn-matdisp}
\end{eqnarray}

\n is obtained where $v_{\rm a}= B_\|/\sqrt{\mu_0\rho}$ is the Alfv\'en speed in the unperturbed field.
This system permits several waves, whose dispersion relation 
takes the form

\begin{equation}
\label{MHDdisp}
 \omega^2 = k^2 v_{\rm a}^2 \pm \zeta v_{\rm sh}^2 \frac{k}{r_{\rm g1}}
\end{equation}

\n similarly to equation~(15) from \cite{Bell2004} where the dimensionless driving parameter $\zeta$ is given by 

\begin{equation}
 \zeta = \mu_0 \frac{j_\|}{B_\|}\frac{v_{\rm a}^2}{v_{\rm sh}^2}r_{\rm g1}.
\label{eqn-driving}
\end{equation}
\n It follows that, for strongly driven modes 
such that $\zeta v_{\rm sh}^2/v_{\rm a}^2 \gg 1$, there exist
aperiodic waves (Re~$\omega=0$), over the range
$1 < k r_{\rm g1} < \zeta v_{\rm sh}^2/v_{\rm a}^2$.
From equation (\ref{MHDdisp}) it is straight forward to show
that the fastest growing mode occurs at

\begin{equation}
k_{\rm max} = \frac{\mu_0}{2} \frac{j_\|}{B_\|}
\label{eqn-kmax}
\end{equation}

\n with growth rate

\begin{equation}
\gamma_{\rm max} = v_{\rm a} k_{\rm max} .
\label{eqn-rate}
\end{equation}

The dispersion relation for arbitrary orientations of $\bfm{k}$, 
$\bfm{j}_{\rm cr}$, and $\bfm{B}_0$ has been determined by
\cite{Bell2005} demonstrating that unstable 
modes exist for all orientations provided $\bfm{k} \cdot \bfm{B}_0\neq 0$.
\cite{melrose} has compared the resonant and non-resonant instabilities
for arbitrary angle of propagation and shown that the latter occur
with elliptically polarised modes.
This emphasises the need for high resolution simulations in three dimensions.

\section{Instability Driven Turbulent Field Generation}
\label{sec-field}

Our simulation box represents a region in the restframe of the upstream magnetised
fluid (shock precursor). We assume a thin box
such that both the pressure and density can be taken to be initially constant,
and the cosmic rays
can be considered unmagnetised (i.e. they travel through the box with straight trajectories). 

\begin{figure}
\centering
\vbox{
   \psfrag{pp1}[Bl][Bl][1.0][+0]{\hspace*{-0cm}$t$}
   \psfrag{pp2}[Bl][Bl][1.0][+0]{\hspace*{-0cm}$B^{\rm rms}_\|$}
   \psfrag{pp3}[Bl][Bl][1.0][-0]{\hspace*{-0cm}$B^{\rm rms}_\perp$}
\includegraphics[width=0.47\textwidth]{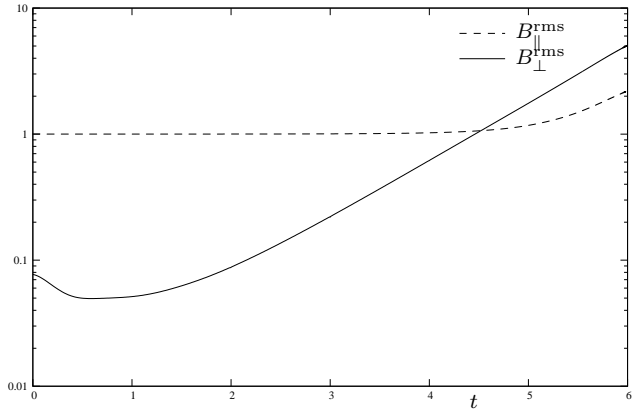}
}
\caption{
The root-mean-square value of
the perpendicular (solid) and parallel (dashed) components
of the magnetic field. The field is initially damped as 
the initial energy input goes into kinetic energy,
setting the inertial plasma into motion. Once the fluid is in
motion the perturbations grow according to the linear analysis 
as $\exp(\gamma_{\rm max} t)$.
}
\label{2ndOgrowth}
\end{figure}

Following~\citet{Bell2004} we take typical interstellar 
values for the upstream quantities $B_\| = 30{\rm n T}$, 
nucleon density $n_0 = 1\times 10^{-6} {\rm m}^{-3}$, corresponding
to an Alfv\'en speed $v_{\rm a}=6.6\times 10^3 {\rm m s}^{-1}$. Furthermore assuming 
$10\%$ energy transfer to cosmic rays and shock speed
$v_s/c=1/30$, with $\ln(p_2/mc)=14$, where $p_2$ is the upper cut-off 
of the cosmic-ray energy spectrum, we obtain 
$\zeta=4.8\times10^{-4}$ and $\zeta v_s^2 = 1100v_{\rm a}^2$.

In the quasilinear theory of diffusive shock acceleration, the
minimum momentum of the cosmic-ray distribution is expected to increase with
distance upstream, since lower energy cosmic rays are confined 
closer to the shock \citep{Eichler79, blasi}.
Assuming Bohm diffusion the minimum cosmic ray momentum at a given distance 
$z$ upstream of the shock is $p_1 \approx 3v_{\rm sh}eBz/c$.  
In order for the non-resonant mode to leave the 
regime of linear growth before being overtaken by the shock front, i.e., 
before being advected over a distance of roughly 
$c r_{\rm g1}/{\rm v}_{\rm sh}$,
we take $p_1 c=1$\,PeV (although this is not a lower limit). 
This gives corresponding physical scales 
$k^{-1}_{\rm max} = 2\times10^{13}\,{\rm m}$, 
$\gamma^{-1}_{\rm max} = 98\,{\rm yr}$, and  
$r_{\rm g1} = 1.1\times 10^{16}\,{\rm m}$. 

\subsection{Turbulent seed field}
\label{sec-turb}

The non-resonant instability develops in a field containing 
a level of seed turbulence. In the simulations that follow
the magnetic field is initialised according to the 
following prescription
\beq
\bfm{B}(x,y,z) = \bfm{B}_0 + \bfm{\delta B}(x,y,z),
\eeq
where ${\bf \delta B}$ is a small isotropic turbulent field component ($\delta B/B_0\ll 1$). The turbulent component is constructed similarly to \cite{GJ94}, by choosing $N$
plane-wave modes with random phases, orientations, and polarization states given by

\begin{equation}
 \delta {\bf B}({\bf r}) = 
2 \sum_{n=1}^N A_n e^{i( {\bfm k}_n \cdot \bfm{r} + \psi_n)} \hat{\xi}_n,
\end{equation}

\n where $\bfm{k}_n\equiv k_n \hat{\bfm{e}}_n^1$. For each mode
$n$, 
$A_n$, $\psi_n$, $k_n$, and $\hat{\bfm{e}}_n^1$ are the 
amplitude, phase, wave-number, and propagation direction respectively. Additionally, the polarization vector $\hat{\xi}_n$ is given by 

\beq
\hat{\xi}_n\equiv\cos(\alpha_n)\hat{\bfm{e}}_n^2+i\sin(\alpha_n)\hat{\bfm{e}}_n^3
\eeq

\n where $\alpha_n$ is the polarization angle. 
The three vectors ($\hat{\bfm{e}}_n^1$, $\hat{\bfm{e}}_n^2$, $\hat{\bfm{e}}_n^3$)
form an orthonormal basis for the plane waves such that, 
for a continuous representation, the field is guaranteed to be 
solenoidal by the condition $\bf{k}_n \cdot \bfm{\xi}_n=0$. 
However, when the field is projected 
onto a mesh of uniform spacing $h$ the solenoidal condition 
on a central-differenced grid is given by \cite{sosdownes},

\begin{equation}
 \hat{\bf \xi}_n\cdot \sin({\bf k}_nh)=0.
\end{equation}

The amplitude of each mode for a given 
variance $\sigma^2$ is 
\begin{equation}
 A_n^2 = \sigma^2 G_n 
\left[ \sum_{n=1}^N G_n\right]^{-1}, \nonumber
\end{equation}
with
\begin{equation}
G_n = \frac{\Delta V_n}{1+(k_nL_c)^\Gamma}\;\;. 
\end{equation}
Here $\Gamma$ is the spectral index of the turbulence, and 
the normalisation factor for three-dimensional turbulence 
is  $\Delta V_n = 4\pi k_n^2\Delta k_n$. $L_c$ is the correlation length of the magnetic field, which in this case we take to be the largest wavelength in
the system.

In order to generate the turbulent field component 4 random numbers are 
required for each mode: two to specify the orientation of the wave-vector
 $\hat{\bfm{e}}_n^1$, and one each for the phase and polarization angles 
$\psi_n$ and $\alpha_n$ respectively. 

Since the turbulent field will be represented on a periodic mesh, each mode must have an integral number of wavelengths in each coordinate direction. Explicitly, assuming a cubic grid of dimension $L$ with $N$ mesh points in each coordinate direction, $\bfm{k} = 2\pi/L (n_x,n_y,n_z)$ where $n_x=0..N/2$ etc. (The factor 2 is necessitated by the hermiticity of $\bfm{\delta B}$ requiring both negative and positive $k$ modes be allowed.) This geometrical constraint also means that at small $k$, the density of unique modes is low and available $\bfm{k}$ vectors will have a strong preferential bias along the grid axes. Since isotropic and homogeneous turbulence is only achieved in the limit of $N\to\infty$~\citep{batchelor}, we do not include modes from the low-$k$ range where this bias is greatest. Additionally, we attenuate the dynamic range at high-$k$ by requiring the shortest scale fluctuations are well resolved over 10 cells (rather than 2 as minimally required above).

\subsection{Numerical Scheme}
\label{sec-num}

We describe the numerical scheme used to perform the simulations of the
non-resonant Bell-type instability. The simulations are performed with
considerably higher resolution than those
performed previously by \cite{Bell2004}. The cosmic-ray current is taken to
be uniform along the $z$-axis and constant in time. 
Kinetic effects are again neglected such 
that ${\bf j}_\perp = 0$. 
For clarity from this point onwards, terms involving magnetic field $\bfm{B}$ are assumed to have absorbed a factor $1/\sqrt{\mu_0}$, and terms involving charge current density $\bfm{j}$ are assumed to have absorbed a factor $\sqrt{\mu_0}$. 
The MHD equations including cosmic ray current may be written in the 
form
\begin{equation}
\label{IVP}
 \pdiff{{\bf U}}{t} + \nabla \cdot {\bf V} = {\bf S},
\end{equation}
where 
\begin{equation}
 {\bf U} = 
(\rho, \rho{\bf u}, {\bf B}, E)^T
\end{equation}
and 
\begin{equation}
 {\bf V} = \left(\begin{array}{c}
 \rho{\bf u} \\
 \rho {\bf uu} + {\bf \mathbb{I}}p^* - {\bf BB} \\
 {\bf uB - Bu} \\
 (E+p^*){\bf u} - {\bf( u \cdot B)B}
\end{array}
\right),
\end{equation}
with $p^*=p+B^2/2$, the total pressure.
The source terms  are given by
\begin{equation}
  {\bf S} = \left(\begin{array}{c}
 0 \\
 -j_\| {\bf \hat{z}} \times {\bf B} \\
 0 \\
 -{\bf u}\cdot(j_\|{\bf \hat{z}}\times{\bf B})
\end{array}
\right).
\end{equation}

We have constructed a finite volume Godunov code in three
dimensions, MENDOZA, which follows the method described in
\cite{FKJ}. The equations are split into three one-dimensional
problems to find the fluxes on the surface of each cell. 
The fluxes are determined at each cell interface
from the solution to the linearised Riemann problem, and the
field is updated with second-order accuracy in both time and space.
The source terms are then added as explicit volume-averaged quantities.
We have investigated a variety of divergence cleaning methods, but found
the generalised Lagrangian multiplier (GLM)
method of \cite{dedner} to be most effective.
Divergence errors are naturally produced in multidimensional
simulations, due to the fact that the one-dimensional problem is solved for
each direction and the fluxes simply added together. 
While the GLM method does not identically conserve $\nabla\cdot {\bf B}=0$,
it has the unique property of rapidly damping any divergence 
introduced, while
advecting them away from the problem areas at the fastest admissible speed.
For the study of magnetic instabilities, particularly in the presence of
strong pressure gradients,
we find this method to have many advantages over correction schemes.
For a comparison of many different divergence cleaning methods see, for
example, \citet{Toth}.

Using simulation units, the model is then initialised with

\begin{center}
\begin{tabular}{llll}
$P = \rho = 1$ , & 
$j_\| = 4\pi/5$ , & 
$B_\| = 1$ , &
$\bfm{u}=0$ . 
\end{tabular} 
\end{center}

\n Equations~(\ref{eqn-driving})~to~(\ref{eqn-rate}) then yield

\begin{center}
\begin{tabular}{lll}
$v_{\rm a} = 1$ , & 
$v_{\rm sh} = 1520$ , &  
$k_{\rm max} = 2\pi/5$ , \\
$\gamma_{\rm max} = 2\pi/5$ ,&
$r_{\rm g1} = 1375/\pi$ . 
\end{tabular} 
\end{center}

As is common practice in the literature, the turbulent field spectrum is represented by a superposition of plane wave modes, one selected in a random direction from each of a number of uniform intervals of $\log k$ over a finite range. As commented in Section~\ref{sec-turb} however, since a discrete mesh is being used to represent a periodic field, viable $k$-vectors are finite in number and may be clustered or multiply degenerate in $k$. Therefore, for a given grid resolution, requiring that each interval contains at least one viable $k$-vector determines the total number of modes used to construct the turbulent field representation and the cutoffs in the power spectrum. In this work, a $512^3$ grid is used with modes in the range $40 \le kL\le 550$. With these parameters
the number of modes we obtain is $N=151$. 
While this dynamic range ($\sim 14$) is relatively small, it does 
represent isotropic, homogeneous turbulence well over the chosen narrow band 
since at least one mode exists within each interval (by construction). 
As long as $k_{\rm max}$ is central to this range, it should make 
little difference to the experiment that the spectrum is narrow since for 
$k>2 k_{\rm max}$ 
the linear growth rate is zero and for $k<2 k_{\rm max}$ the growth 
rate goes as $k^{1/2}$~\citep{Bell2004}. We choose $L=64$ such that 
$k_{\rm max}L = 80$ and the corresponding fastest-growing fluctuation 
scale is well resolved over $\sim 40$ cells.

In simulation units, 
the initial seed field is set with $\sigma^{2} = 0.01 B_\|^2$. Finally, we restrict our studies to the spectral index corresponding to three-dimensional Kolmogorov turbulence, $\Gamma = 11/3$, \citep[e.g.][]{GJ99}.

\begin{figure}
\centering
\vbox{
  
\includegraphics[width=0.45\textwidth]{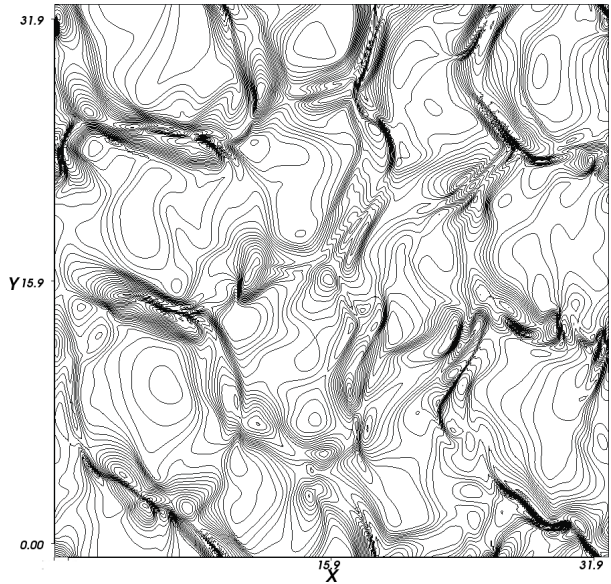}
}
\caption{
Slice through the $x$-$y$ plane at the centre of the simulation box
at $t=6$ showing the isolines of the magnetic
field magnitude. 
Contours have uniform spacing of $0.5$ 
with $B_{\rm min}=0.5$ and $B_{\rm max}=13$.
The regions of strongest field are found in the regions of higher density.
}
\label{cav_fig}
\end{figure}

\subsection{Results and Discussion I}
\label{sec-res1}

Initialising the fields using the parameters described in sections
\ref{sec-turb} and \ref{sec-num}, the development of the 
root mean squared parallel and perpendicular components of the 
magnetic field is plotted in Figure~\ref{2ndOgrowth}. Since the plasma 
starts from rest, the energy input at the beginning primarily goes
into kinetic energy of the background fluid and no field growth is observed. 
After this brief initial stage, at approximately $t=1$, the
magnetic field starts to grow at close to the linear growth rate 
$\gamma_{\rm max}$.
The energy in the perpendicular magnetic field eventually overtakes
that of the parallel field but continues to grow at the same rate.
As the amplitude of the unstable modes approaches that of the 
mean field, cavity structures similar to those observed in the simulations 
of \cite{Bell2004} start to appear. The development of these 
cavities is driven by the Lorentz force exerted on 
the circular/elliptical modes 
by the return current. The expansion of each cavity
is hindered by the growth of neighbouring cavities and 
dense regions are formed between them. The frozen-in magnetic field 
becomes very large inside these cavity walls, as shown
in Figure~\ref{cav_fig}.

It is at this point that neglecting the non-linear feedback of the
field on the cosmic rays becomes apparent. As discussed in
\cite{Bell2005}, information is only passed along the direction 
of the cosmic ray current via tension in the magnetic field lines.
However, as the field increases, the cosmic rays themselves
would most likely be beamed into the resulting cavities. From the
simulations, it is observed that there are 
no clear coherent structures forming along 
the direction of the cosmic ray current. 

A second difficulty is that vacuum states begin to form 
within the cavities, and the MHD equations lose their validity. A
possible solution to this is to use a highly dissipative scheme,
but this reduces the accuracy of the solution. Using such a method
the results obtained
for the long term evolution of the field are similar to those in
\cite{Bell2004}, and we do not comment on them.

\begin{figure}
\centering
\vbox{
   \psfrag{pp1}[Bl][Bl][1.0][+0]{\hspace*{-0cm}$kL$}
   \psfrag{pp2}[Bl][Bl][1.0][-0]{\hspace*{-0cm}P(k)}

   \psfrag{pp3}[Bl][Bl][1.0][-0]{\hspace*{-1cm}$t=5$}
   \psfrag{pp4}[Bl][Bl][1.0][-0]{\hspace*{-1cm}$t=6$}
 
   \includegraphics[width=0.32\textwidth, angle=-90]{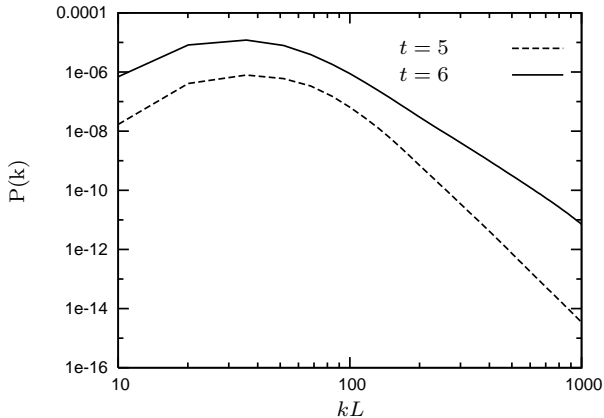}
}
\caption[Power spectrum of amplified field]{
Logarithmic plot of the power spectrum in the amplified field at
times $t=5$ and $t=6$. The spectrum at large $k$ is a power law with 
$P(k)=k^{-s}$, with $s\approx 7.5$ at $t=5$ and $s\approx 5$ at $t=6$.
The turnover in the spectrum moves to smaller $k$ as the 
parallel component of the field increases.
}
\label{MendozaPower}
\end{figure}

Since we are primarily concerned with determining diffusion statistics
in an amplified field we choose to take a snapshot of the field in the
early stages of the non-linear development, at $t=6$.
The power spectrum of the field at $t=5$ and $t=6$ is plotted in 
Figure~\ref{MendozaPower}. 

In the shock frame, the amplified field is advected towards the shock.
It is expected that lower energy particles are confined closer to the shock 
by the turbulence produced further upstream by the cosmic rays which can 
diffuse further from the shock.
We determine the transport properties for relativistic protons in a
static amplified field taken at $t=6$.

\section{Cosmic Ray and Field Line Transport in Turbulent Field}
\label{sec-part}

We now discuss the first results of cosmic-ray transport in a turbulent field resulting from the non-resonant instability of \cite{Bell2004} as described in Section~\ref{sec-inst}. In particular, we investigate how the transport properties
differ from the Bohm diffusion
limit. Much work has been done on this topic numerically by 
~\cite[][for example]{GJ94, GJ99, Casseetal} but in all cases an \emph{a priori} field was assumed. However, the field we use has been 
determined self-consistently.   

\begin{figure}
\begin{center}
\includegraphics[width=0.47\textwidth]{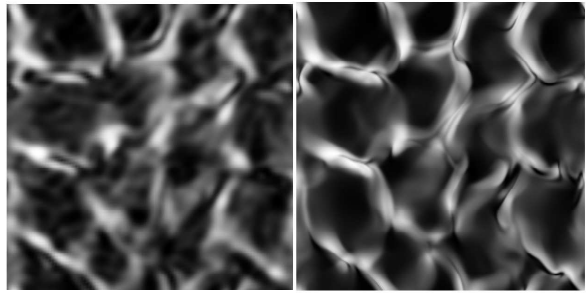}
\caption[Comparison of FFT field with MENDOZA output]{
A comparison of the fields after being passed through the low
power filter. The same large scale structure maintained
quite well between the filtered field (left) and MENDOZA's
field (right).
}
\label{FFTcompare}
\end{center}
\end{figure}

To determine the 
statistical transport properties of the amplified field we have developed the code
\lq\lq LYRA\rq\rq~\citep{OSDBB08}. LYRA 
assumes a Fourier mode description of a static magnetic field as described in Section~\ref{sec-turb}. However, MENDOZA represents the magnetic field at a finite number of mesh points (in our case, at $512^3$ uniformly spaced points). We convert the field into a continuous Fourier mode representation by taking a Fourier transform and importing the 5000 highest power real modes into LYRA. In the particular case considered here, passing the field through what amounts to a low-power filter results in a loss of $16\%$ of the power in the field. Any lost power is returned to the field by rescaling the included modes accordingly. Figure~\ref{FFTcompare} provides a comparative illustration of the pre-processed field generated by MENDOZA and the filtered field read in by LYRA for the particle transport simulations. 
As can be seen, the overall large scale structure
of the field is maintained quite well, but there is a certain amount of smoothing 
of the sharper features. However,
the continuous description of the field provided by the Fourier 
transform is more practical than a piecewise constant mesh for the purposes of
integrating particle trajectories. The resulting field is also divergence free. 
We can not be certain that the spreading of the field inside 
the cavity walls does not affect the final results, but in the current work
the Fourier description makes the computations much simpler.
We will address this issue in future work.

Another important limitation introduced by importing the 
field from a periodic grid
representation is that while, in theory, we have access to an arbitrarily large
dynamic range, the field will remain periodic on the relatively small scale of 
the parent field. It is important to consider this point when interpreting 
results displaying any similarity on scales corresponding to the periodicity 
of the field. 

With the field described in this manner, the equations of motion  
are integrated using the Burlisch-Stoer method~\citep{NumericalRecipes}. 
In the case of field lines, the equation to be integrated is
$\mbox{d}\bfm{r}/\mbox{d}s= \bfm{B}/B$. For protons, the governing 
equations are

\begin{equation}
\dif{\gamma {\bf v}}{t}=e{\bf v \times B} ,
\end{equation}

\n and

\begin{equation}
\dif{{\bf r}}{t}={\bf v} .
\end{equation}

\n When integrating the equations of motion for protons, the
tolerances in the integrator are set to conserve energy to  $\lesim 1\%$ over $1000/2\pi \omega_{\rm g}^{-1}$ where $\omega_{\rm g}^{-1}\equiv 2\pi r_{\rm g}$.  
We determine the statistical transport properties 
for 9 different energy values
ranging from $\gamma=10^3$ to $\gamma=4\times10^5$.
For each value, an ensemble of 2000 particles is 
initialised with random starting
positions and pitch angles.
Using the $t=6$ field snapshot described in the previous section, 
with magnetic field simulation units expressed in terms of 
a $3\mu$G ISM field value,
we carry out particle transport simulations using LYRA.

\subsection{Results and Discussion II}
\label{sec-res2}

\begin{figure}
\centering
\vbox{
   \psfrag{pp1}[Bl][Bl][1.0][+0]{\hspace*{-0cm}$D_{\perp}$}
   \psfrag{pp2}[Bl][Bl][1.0][-0]{\hspace*{-0cm}$s$}
\includegraphics[width=0.43\textwidth]{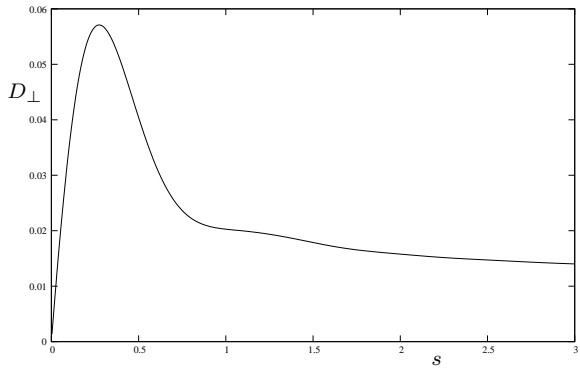}
}
\caption[Perpendicular magnetic diffusivity]{
The Perpendicular magnetic diffusivity plotted against distance
along a field line. Both in in units of box lengths.
}

\label{field_diff_fig}
\end{figure}

To investigate the properties of the field lines themselves
2000 field lines are integrated, and the mean square perpendicular
displacement along each field line is calculated. 
The magnetic diffusivity $D \equiv \langle\Delta x^2\rangle/2s$,
where $s$ is the distance along a field line, is plotted in 
figure~\ref{field_diff_fig}. Since the fastest growing modes are
circularly polarised, the field lines are helical, although the
guiding centre wanders about the $x$-$y$ plane as it moves along the 
$z$-axis. The parallel diffusivity is essentially ballistic 
($D_{\|} \propto s$).

We numerically determine the statistical average diameter of the
helices by iterating over the same 2000 field lines, and approximating
the diameter of each revolution. In what follows, all length units are
normalised to simulation box lengths.
The root mean square diameter of the helices is $d=0.148$ which is
an important length scale 
in the study of the fields properties. 
The peak in the perpendicular magnetic diffusivity is located at
$s=0.273$. Using this value and setting the perpendicular displacement 
equal to the root mean square
diameter gives $D_\perp=0.08$ which is in reasonable agreement with the experimentally
determined peak at approximately $D_\perp=0.057$,
as shown in figure~\ref{field_diff_fig}.
\begin{figure}
\centering
\vbox{
   \psfrag{pp1}[Bl][Bl][1.0][+0]{\hspace*{-0cm}$t$}
   \psfrag{pp2}[Bl][Bl][1.0][-0]{\hspace*{-0cm}
$\frac{\langle \Delta z^2\rangle}{2 \Delta t}$}

\includegraphics[width=0.32\textwidth, angle=-90]{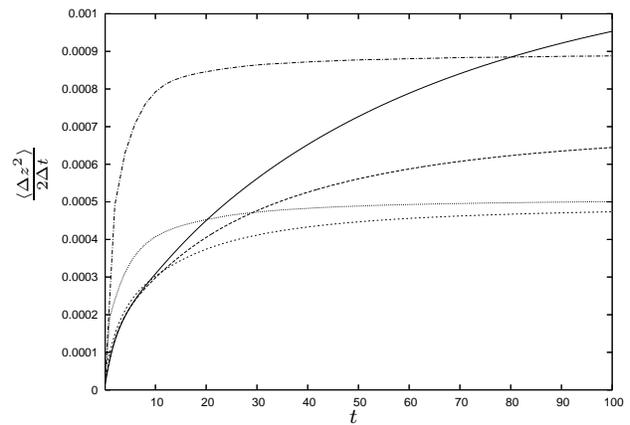}
}
\caption{
$\frac{\langle \Delta z^2\rangle}{2 \Delta t}$ at t=6 showing times for super-diffusive regime in units
of box-lengths squared per Larmor time. 
Lines correspond to gamma factors (from bottom to top at $\Delta t=100$)
$\gamma = 4\times10^3,~10^4,~2\times10^3,~2\times10^4,~10^3$.
Time units are Larmor times for a $\gamma=10^4$ particle. Note these
lines continue to asymptotic values and we show only the early
evolution to demonstrate the super-diffusive behaviour.}
\label{t6pll}

\end{figure}

The mode with the fastest growth rate has a wavelength similar
to the gyroradius of protons with a particle gamma factor of 
$\gamma\approx2\times10^4$. We focus on the transport properties
close to this value. Particles with $\gamma$ much less or greater than this 
have mean free paths close to a cell width or box length respectively.

We interpret these results in terms of 
the diffusion coefficients for the actual particle
trajectories determined from LYRA. The parallel and perpendicular diffusion 
coefficients are determined by averaging over a large ensemble of particles
\begin{equation}
\kappa_\| \equiv \lim_{\Delta t \rightarrow \infty}
\frac{\langle \Delta z^2\rangle}{2 \Delta t}
\end{equation}
\begin{equation}
\label{kappa_perp}
\kappa_\bot \equiv \lim_{\Delta t \rightarrow \infty}
\frac{\langle \Delta x_\bot^2\rangle}{2 \Delta t}
\end{equation}

\n Note that the $\kappa_\|$ and $\kappa_\perp$ refer to
parallel and perpendicular directions with respect to 
the initial field, whereas in the $t=6$ snapshot of the field
the perpendicular component already dominates, as is evident from 
Figure~\ref{2ndOgrowth}. 

\begin{figure}
\centering
\vbox{
   \psfrag{pp1}[Bl][Bl][1.0][+0]{\hspace*{-0cm}$t$}
   \psfrag{pp2}[Bl][Bl][1.0][-0]{\hspace*{-0cm}
$\frac{\langle \Delta x_\bot^2\rangle}{2 \Delta t}$}

\includegraphics[width=0.32\textwidth, angle=-90]{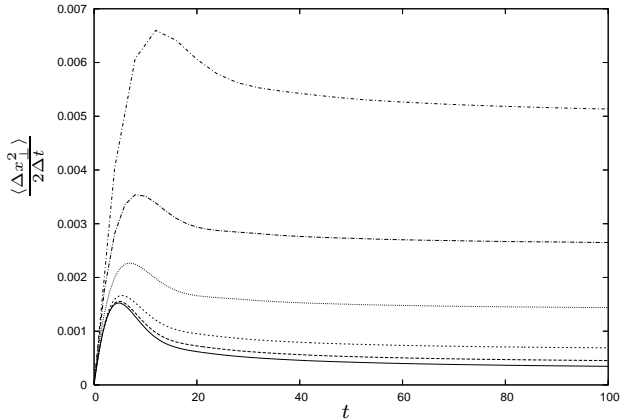}
}
\caption{
$\kappa_{\perp}$ at $t=6$ showing times for super-diffusive regime in units
of box-lengths squared per Larmor time. 
Lines correspond to gamma factors (from bottom to top)
$\gamma = 10^3,~2\times10^3,~4\times10^3,~10^4,~2\times10^4,~4\times10^4$.
Time units are Larmor times for a $\gamma=10^4$ particle .}
\label{t6perp}
\end{figure}

The initial evolution of the particle trajectories are 
governed by the field line statistics. Figure~\ref{t6pll}
and Figure~\ref{t6perp} plot the early parallel
and perpendicular diffusion respectively, 
showing an initially super diffusive regime, with a peak
in the perpendicular value for the lower energy particles 
occurring at approximately 5 $\omega_{\rm g}^{-1}$. Inserting this values into Eq.~(\ref{kappa_perp})
and equating to the experimental peak value $\sim 0.0015$, gives an estimate for the
root mean square perpendicular displacement at the peak, 
$\sqrt{\langle \Delta x_\bot^2\rangle} = 0.08$. This value is also in agreement with
the statistical mean radius of the helices. For higher energy particles the Larmor radius of a particle becomes
comparable to or larger than the average diameter, 
and the peak time differs from this value.
These particles are not trapped to individual field lines
and must travel a greater distance before becoming diffusive.

\begin{figure}
\centering
\vbox{
   \psfrag{pp1}[Bl][Bl][1.0][+0]{\hspace*{-0cm}$\kappa_\|/\kappa_{\rm Bohm}$}
   \psfrag{pp3}[Bl][Bl][1.0][-0]{\hspace*{-0cm}$\kappa_\perp/\kappa_{\rm Bohm}$}
   \psfrag{pp2}[Bl][Bl][1.0][-0]{\hspace*{-0cm}$\gamma$}
\includegraphics[width=0.47\textwidth]{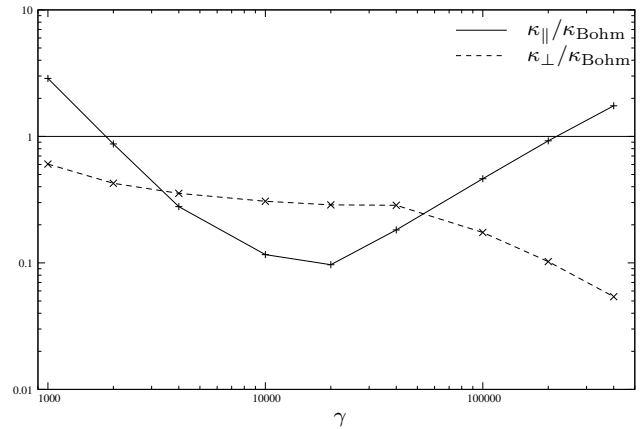}
}
\caption{
$\kappa_\|/\kappa_{\rm Bohm}$ (solid) and 
$\kappa_\bot/\kappa_{\rm Bohm}$ (dashed) with 
$\kappa_{\rm Bohm}$ evaluated in a $3\mu{\rm G}$ field,
at t=6.}
\label{koverbohm}
\end{figure}

The asymptotic values of the diffusion coefficients, after the non-diffusive 
regimes, are plotted in figure~\ref{koverbohm}. It is clear that 
$\kappa_\| < \kappa_\bot < \kappa_{\rm Bohm}$, in the energy range
corresponding to $k_{\rm max}r_{\rm g} \approx 1$. 
In a uniform field, quasi-linear theory predicts
$\kappa_\bot < \kappa_{\rm Bohm}< \kappa_\| $. However, 
in the highly non-uniform field produced by the non-resonant
instability, the experimentally determined transport properties are 
much more complicated.
Nevertheless, the parallel diffusion coefficient, along with the shock speed, 
defines the residence and acceleration timescales. While the simulations presented in this paper do not include resonant wave excitation, 
which would act to reduce the diffusion coefficients even further, 
it is worthwhile speculating on the effects that our results 
would have on the shape of the CR spectrum and the rate of the shock acceleration 
process. The acceleration rate defined by \cite{LagageCesarsky} 
is proportional to the magnetic field strength. Amplification of the magnetic 
field, as in the snapshot we selected, above interstellar medium values 
would therefore increase the maximum energy limit given by \cite{LagageCesarsky}. 
Inclusion of resonant wave excitation and application to the 
case of a circumstellar wind would increase that 
limit even further \citep{belllucek} as, of course, would a higher
saturation level of the non-resonant waves.

Turning to the shape of the spectrum, at a strong shock 
front with a compression ratio of four the standard result that $f\propto p^{-4}$ applies provided 
that particles are isotropised and that their motion becomes diffusive within a residence 
time either side of the shock \citep{duffyetal}. In our simulations the isotropisation is very 
rapid, on the timescale of one to two Larmor times. However, the parallel motion only becomes 
fully diffusive after almost ten Larmor times and is super-diffusive ($<\Delta x^2>\propto t^\beta$ 
where $\beta>1$) up to that point. It has been shown \citep{kirketal} that if particle crossings
happen on a timescale below that required for diffusive transport then the 
resulting spectrum will be a power law given by $f\propto p^{-(5-\beta)}$ 
provided the distribution remains close to 
isotropy. In our super-diffusive case, $\beta>1$, 
this would imply a spectrum that is harder than 
the diffusive shock acceleration result. Therefore, 
our results are consistent with the rapid 
acceleration of a relatively hard spectrum,but clearly more 
work needs to be done on resonant wave excitation and overall dynamical range before drawing definitive conclusions.


\section{Conclusions}
\label{sec-conc}

We have reported on the results from high-resolution MHD simulations 
of the non-resonant current driven instability. The long term evolution of
the field is still uncertain, as the non-linear development leads to 
very low density cavities in the absence of significant numerical dissipation. 
To overcome this, future simulations should include a non-linear feedback 
on the cosmic rays perhaps using a hybrid code, 
similar to the work of \cite{lucekbell}, or by performing   
large-scale particle in cell (PIC) simulations. 
To this end, \cite{Riquelme} have performed 2D PIC 
simulations of the relativistic generalisation
of the Bell-type instability~\citep{revilleetal06} and
preliminary results show that the field
saturates with both the perpendicular and parallel magnetic 
field well in excess of the initial mean value. 
On the other hand, \cite{pohlICRC} have reported on 
3D PIC simulations, with initial parameters chosen such that 
the non-resonant mode is expected to be observed. 
However, the filamentation instability dominates the growth of
the field and the perturbations do not grow much beyond the 
initial mean field value. The results of PIC simulations
of this instability remain inconclusive.
The simulations of \cite{Riquelme} are promising in that they 
do show the development of extended filaments, 
as predicted in \cite{Bell2005}, and this should be 
important for future investigations of particle transport.

In recent years, magnetic field amplification has become a popular 
mechanism for accelerating cosmic rays beyond the Lagage-Cesarsky limit.
We present here, the first attempt at investigating 
self-consistently the particle diffusion statistics
in a self-excited magnetic field. The observed diffusion is anisotropic
($\kappa_\| \neq \kappa_\perp$), with the low energy particles' diffusion
being largely determined by the field statistics. Thus at low energies
$\kappa_\| > \kappa_\perp$. However, on length scales similar to the
wavelength of the helical modes, motion parallel to the shock normal
is more diffusive.
At high energies the diffusion seems to be energy independent 
perpendicular to the initial field direction, but scales as $\gamma^2$ 
parallel. This follows naturally since there are few waves with
scalelength comparable to the gyroradii of these particles in our 
simulation box. However, the simulations do achieve sub-Bohm diffusion 
with only modest amplification of the effective ambient field.

It is clear that in the presence of non-linear waves
the diffusion properties of particles differ greatly from the
standard picture.
Based on the results presented here it is difficult to say if
the super-diffusive or sub-diffusive regimes will have a significant
effect on the shape of the spectrum. An estimation of the acceleration
timescale of cosmic rays is beyond the scope of this paper, 
nevertheless we have demonstrated that self-excited turbulence does
reduce the diffusion coefficient of the cosmic rays and is therefore
likely to lead to acceleration beyond the Lagage-Cesarsky limit.

\section*{Acknowledgments}

This work was partly funded by the CosmoGrid project, funded under the
Programme for Research in Third Level Institutions (PRTLI) administered by the
Irish Higher Education Authority under the National Development Plan and with
partial support from the European Regional Development Fund.
The authors wish to acknowledge the SFI/HEA Irish Centre for High-End Computing (ICHEC) for the provision of computational facilities and support.
BR and SOS thank D. Brodigan and D. Golden for technical support 
on the ROWAN computing facilities. BR acknowledges helpful
discussions with A.R. Bell. 
PD would like to thank the Max Planck Institute for Nuclear Physics
in Heidelberg for their hospitality. PD thanks the Dublin Institute for Advanced Studies
for hosting him during 2006/2007.

\label{lastpage}

\end{document}